Thermally driven Josephson oscillations in superfluid $^4$He

E. Hoskinson and R. E. Packard

Department of Physics, University of California, Berkeley, CA 94720-7300

We find that a temperature differential can drive superfluid oscillations in $^4$He. The oscillations are excited by a heater which causes a time dependent temperature differential across an array of 70nm apertures. By measuring the oscillation frequency and simultaneously determining both temperature and pressure differentials we prove the validity of the most general form of the Josephson frequency relation. These observations were made near saturated vapor pressure, within a few mK of the superfluid transition temperature.

Macroscopic quantum systems such as superconductors, superfluids and Bose Einstein condensates are often described by a complex order parameter, $\psi = \rho_s e^{i\phi}$. Free energy gradients are equivalent to unbalanced forces. When such gradients exist the dynamics of the medium can be described by the time evolution of the phase difference between two points. The general equation of motion is given by the Josephson-Anderson phase evolution equation [1],

$$\frac{d\Delta\phi}{dt} = -\frac{\Delta\mu}{\hbar} \qquad \textbf{1,}$$

where $\Delta$ signifies the difference of the quantity between two points, $\mu$ is the chemical potential, and $\hbar$ is Planck's constant $h$ divided by $2\pi$. Perhaps the greatest utility [2] of Eq. 1 arises when the phase difference changes periodically in units of $2\pi$ for example in weak links characterized either by a Josephson-like sinusoidal current-phase relation, $I(\phi) \propto \sin(\phi)$, or a linear current-phase relation with vortex-like phase slippage. When $\Delta\mu$ is constant in time, the frequency of $2\pi$ cycles is given by the Josephson frequency formula,

$$f_j = \Delta\mu/h \qquad \textbf{2}.$$

In superconductors the dominant chemical potential gradient arises from impressed voltages. Until now the chemical potential drive in superfluids has been provided by an externally imposed pressure gradient. However, more generally, in a neutral fluid the complete form of chemical potential involves both pressure and temperature [3]:

$$\Delta\mu = m(\Delta P/\rho - s\Delta T) \qquad \textbf{3},$$

where m is the particle mass, $\rho$ is the fluid mass density and s is the entropy per unit mass.

When two superfluid samples are connected by an array of small apertures, Josephson oscillations driven solely by pressure differentials have been shown to be in quantitative agreement with Eq. 2. for both superfluid $^3$He [4] and $^4$He [5]. In this paper we report the first observation of superfluid Josephson oscillations driven by a temperature differential, thereby demonstrating the most general form of the Josephson frequency relation.

We focus here on $^4$He driven through an aperture array using the apparatus schematically shown in figure 1. This is similar to previous experimental cells used in our laboratory except now we have included a heater inside the small inner cell. The array is 65x65 nominally 70nm apertures spaced on a 3μm square lattice in a 50nm thick silicon-nitride membrane.

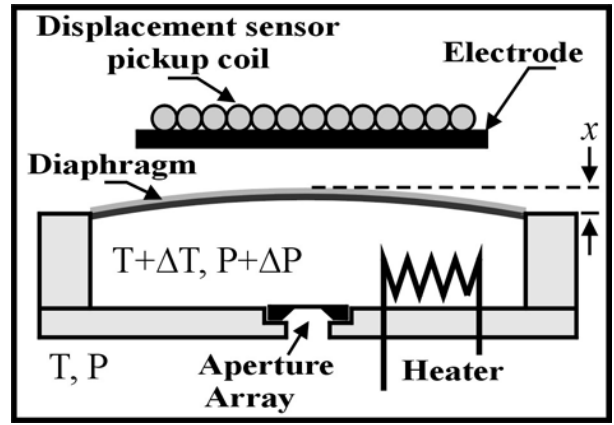

**Figure 1.** Schematic of the experimental cell. A cylindrical washer-shaped spacer of height 0.6mm and inner diameter 8mm is bounded on the top by a flexible Kapton diaphragm and on the bottom by a rigid plate containing the aperture array. The array is produced using e-beam lithography in a 50nm thick 200μm x 200μm silicon nitride membrane supported by a silicon frame that is glued into the lower plate. The top surface of the Kapton is coated with lead. Its position, determined by an adjacent superconducting displacement sensor [6] is proportional to $\Delta P$ and acts as a pressure gauge with resolution better than 1 μPa/Hz$^{1/2}$. The velocity of the diaphragm is proportional to the total mass current $I_t$ flowing through the aperture array. The heater inside the inner cell is a 54mΩ length of CuNi wire, flattened and roughened to increase surface area, to which electrical leads are attached. Superconducting NbTi wire (50μm diameter) is used for the leads to minimize thermal conduction along them and ensure all power delivered is deposited inside the inner cell. The cell sits inside a metal can which is immersed in a pumped bath dewar of liquid helium. The can and cell are filled with $^4$He through a cryogenic valve to a nominally zero ambient pressure. The temperature of the bath and the $^4$He inside the can is controlled by a standard feedback loop.

When heat is applied, the temperature inside the inner cell begins to rise, creating a temperature difference $\Delta T$ across the aperture array. Josephson oscillations are observed, beginning at a low frequency which increases quickly with $\Delta T$. A net dc super-current into the inner cell, expected with the 2π phase slip oscillation mechanism, causes a pressure $\Delta P$ to build, counteracting $\Delta T$ in the chemical potential difference. The Josephson frequency reaches a peak and drops again as the $\Delta P$ term catches up to the $\Delta T$ term. Eventually $\Delta P$ reaches the fountain pressure $\Delta P_f = \rho s \Delta T$, where $\Delta \mu = 0$, and the Josephson oscillations cease. The final steady-state value of $\Delta T$ is determined by a balance between heat introduced, by the heater, and heat lost, by conduction through the inner cell walls (mainly) and normal component flow out through the apertures (a small contribution).

We directly determine the pressure difference $\Delta P$ from the diaphragm displacement transducer. Our main goal here is to determine $\Delta T(t)$ in order to test the generalized Josephson frequency relation, Eq. 2, including the full expression for $\Delta \mu$, Eq. 3.

The inner cell temperature increase, $\Delta T(t)$, is determined by the balance of four heat flows in the cell. The power applied to the heater causes a temperature increase. Superfluid current $I_s$ flowing into the inner cell causes the inner temperature to drop (the thermo-mechanical effect). Normal current flowing out of the cell, $-I_n$, carries heat with it, also causing cooling. Finally, thermal conduction through the walls of the inner cell acts to reduce $\Delta T$. We show in Appendix A that $\Delta T$ evolves according to

$$C_p \frac{d\Delta T}{dt} = -sT(I_s - \frac{\rho_s}{\rho_n} I_n) - \frac{\Delta T}{R} + W_h. \qquad 4.$$

Here $\rho_s$ and $\rho_n$ are the superfluid and normal fluid densities, $W_h$ is the power applied to the heater, $R$ is the thermal resistance between the $^4$He inside the inner cell and the $^4$He outside it, and $C_p = c_p V$, where $c_p$ is the heat capacity per unit volume of $^4$He and $V$ is the volume of the inner cell.

The size of the apertures is such that the viscous normal flow, while small in comparison with the superflow, is not entirely negligible. Normal flow obeys a Navier-Stokes equation with the addition of a $\nabla T$ term [7]. With this $\nabla T$ term, flow through a constriction takes the form

$$I_n = -\frac{\rho_n \beta}{\eta}(\rho_n \Delta P/\rho + \rho_s s \Delta T) \qquad 5,$$

where $\beta$ is a geometrical factor and $\eta$ is the viscosity. Note that if $\Delta \mu = 0$, the quantity in brackets is simply $\Delta P_f$. We are able to measure $\beta$ directly from the flow response to an electrostatically impressed $\Delta P$ just above the superfluid transition temperature $T_\lambda$, where $\rho_s = 0$ and $\rho_n = \rho$.

The superfluid current is determined from the measured total current $I_t = I_s + I_n = \rho A \, dx/dt$ by subtracting the normal current. Here $x$ is the displacement of the diaphragm, whose area is $A$.

The volume of $^4$He outside the inner cell is large enough that its temperature $T$ can be taken to be constant. We measure the bath temperature $T$ using a carbon resistance thermometer calibrated against the vapor pressure. We identify $T_\lambda$ within the cell by detecting a change in the response of the diaphragm when it is electrostatically driven. Calibration of $\Delta P$ and $I_t$, and measurement of $R$ and $V$ is described in Appendix B. Using these parameters and published values for $s$, $\rho_s$, $\rho_n$, $\eta$, and $c_p$ [8], we numerically integrate Eq. 4 to determine $\Delta T(t)$.

A typical transient is shown in figure 2. Here we display the values of $m_4 \Delta P/\rho$ and $m_4 s \Delta T$ as well as the complete chemical potential. As described above, one can see the temperature increase quickly, the pressure rise to meet it, and a state of zero chemical potential difference finally attained.

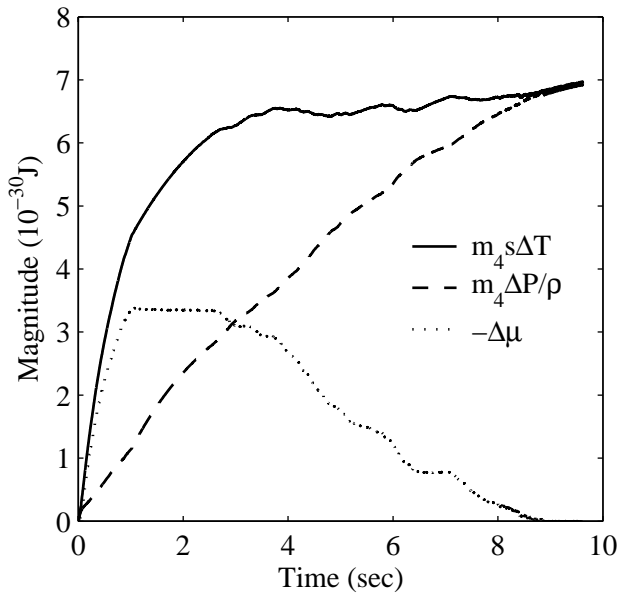

**Figure 2**. A typical thermally driven transient. At time t = 0, the heater is turned on, delivering a constant power to the inner cell. The figure shows the evolution of the temperature term $m_4 s \Delta T$, pressure term $m_4 \Delta P / \rho$, and chemical potential $\Delta \mu$ across the aperture array. The maximum magnitude $7 \times 10^{-30}$ J corresponds to $\Delta T = 0.67\,\mu K$ and $\Delta P = 0.15$ Pa. The pressure $\Delta P(t)$ is directly measured, while $\Delta T(t)$ is determined from the competing heat flows. This particular transient was taken at $T_\lambda - T = 1.5$ mK, a regime where the flow transients last sufficiently long to clearly identify the oscillation frequency and its variation with time.
\

When $\Delta \mu \neq 0$, we detect Josephson oscillations on the diaphragm. We divide the transient into a sequence of small time intervals and determine the frequency in each interval from a Fourier transform. We plot this frequency against $\Delta \mu$ in Figure 3. The figure strikingly shows that although each of the two competing chemical potential terms varies appreciably over the transient, the Josephson frequency is precisely proportional to the entire chemical potential difference even when $\Delta \mu$ is determined in large part by $\Delta T$. Furthermore within the experimental accuracy, the inverse slope of the curve is Planck's constant, in complete agreement with the Josephson frequency formula. We believe this is the first demonstration of Josephson oscillations arising from the temperature term in the chemical potential.

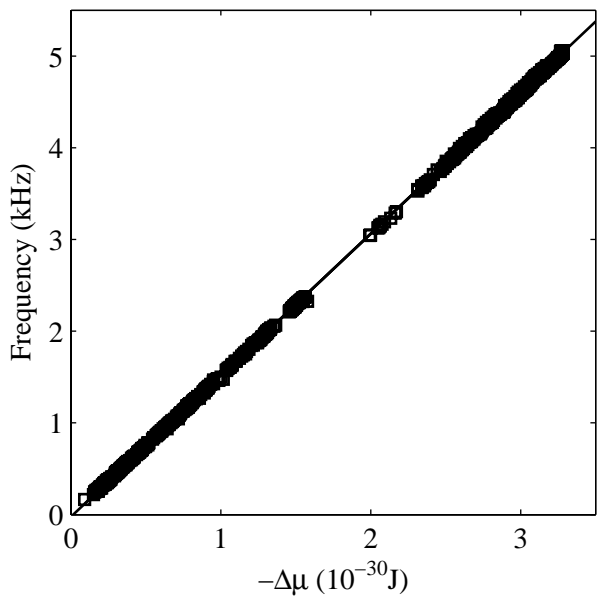

**Figure 3**. Demonstration of the generalized Josephson frequency relation. The observed oscillation frequency is plotted against the full chemical potential difference for the transient shown in figure 2. A straight line least-squares fit (solid line) to the data (squares) gives an inverse slope of $(1.02 \pm 0.02)h$, where $h$ is Plank's constant, and an offset of $-0.02 \pm 0.03 \,\text{kHz}$. The amplitude of the oscillation varies considerably with frequency due to interactions with acoustic resonances of the cell. For most of the transient, the oscillation is clearly discernable and the frequency can be determined with negligible uncertainty. The gaps in frequency data correspond to intervals in which the oscillation amplitude has dropped below the background noise.

In addition to a 2% uncertainty in our pressure calibration, the following sources contribute to the uncertainty in the fitted slope of figure 3.

The carbon resistance thermometer used has a resolution on the order of $10\mu\text{K}$ for a ten second integration interval. The location of $T_\lambda$, however, was determined to within only 0.2mK. This translates into an uncertainty in the values used for $c_p$ and $\rho_s$ of 1% and 10%, respectively, and negligible uncertainties in $s$, $\rho_n$, and $\eta$. The relatively large uncertainty in $\rho_s$, however, does not affect the uncertainty in the calculated $\Delta T(t)$ or $\Delta \mu(t)$ because the terms with $\rho_s$ (in equations 4 and 5) are negligible at $T_\lambda - T = 1.5\text{mK}$.

The bath temperature $T$ must be quite stable during a transient measurement. Any drift can give rise to a $\Delta T_{drift}$ and associated fountain pressure $\Delta P_{drift}$ across the aperture array. The large heat capacity of the bath ensures that only slow drifts occur, which are minimized by feedback control. We are able to measure $\Delta P_{drift}$, and have determined that the resulting error in $\Delta T(t)$ and $\Delta P(t)$ in figure 2 is no more than 4%. The $T$ drift error in $\Delta \mu(t)$ is actually considerably less, since for these small slow drifts, $\Delta \mu_{drift}/m_4 = \Delta P_{drift}/\rho - s\Delta T_{drift} \approx 0$.

The above sources of error combine to give a maximum 4% uncertainty in the calculated chemical potential difference, which in turn yields the 2% uncertainty in the fitted slope of figure 3.

This paper does not address the interesting question of the precise current-phase relation of the weak link array. The criterion for a $I \propto \sin(\phi)$ weak link is that the healing length be larger than the aperture dimensions. For $^3$He the healing length $\xi_3$ is comparable to the aperture dimensions and the array is indeed characterized by a sinusoidal current-phase relation [9]. The healing length for $^4$He varies near the superfluid transition $T_\lambda$ as, $\xi_4 = \xi_o/(1 - T/T_\lambda)^\nu$, where $\xi_o \approx 0.3 nm$, $T_\lambda = 2.176\,\text{K}$ and $\nu = 0.67$, although estimates of $\xi_o$ as low as 0.1nm have been used. We therefore believe that for the transient data like that shown in Figures 2 and 3, taken a few mK below $T_\lambda$, the healing length is smaller than the size of the apertures, the current-phase relation $I(\phi)$ is rather linear, and the oscillation is due to dissipative phase slips. One expects that for temperatures closer to $T_\lambda$ the apertures might act like ideal weak links as in the $^3$He case and exhibit a sinusoidal current-phase relation. Indeed, flow features in a hydrodynamic resonator at $T_\lambda - T \leq 60\mu K$ have been reported to be consistent with a sine-like current phase relation [10] in $^4$He.

In this work we were mainly limited to the temperature regime a few mK below the transition temperature $T_\lambda$ for two reasons. Measurements much closer to $T_\lambda$ require better temperature stability than we have at present. At temperatures much below $T_\lambda$ the superfluid critical velocity is large [11], such that the transients occur in a short time, and the oscillation frequency changes so rapidly that we cannot discern discrete frequencies. We have, however, observed the oscillations as low as 150mK below $T_\lambda$ and they do not appear to be different in nature from the data presented here. Determination of the exact form of the current phase relation, and the expected cross-over to ideal weak-link sinusoidal behavior, remains an intriguing problem.

In conclusion we have experimentally demonstrated that a temperature difference drives Josephson oscillations in $^4$He, in accordance with the generalized Josephson frequency formula. These results may lead to the development of a dc chemical potential drive based on temperature differences alone, that could be used in the implementation of a sensitive superfluid gyroscope [12,13].

**Appendix A – Derivation of $d\Delta T/dt$ equation.**

The equation for $d\Delta T/dt$ can be derived by considering the rate of change of the total inner cell entropy, $dS/dt$. Superfluid flow carries no entropy, whereas normal flow into the inner cell carries it at a rate $s(\rho/\rho_n)I_n$. The heater causes $S$ to increase at a rate $W_h/T$, and conduction through the walls causes it to decrease at $\Delta T/RT$. Combining these three contributions gives $dS/dt = s(\rho/\rho_n)I_n + W_h/T - \Delta T/RT$. Taking $T$, $P$, and $N$ to be the independent variables (where $N$ is the number of $^4$He atoms in the inner cell, so that $I_t = m_4\, dN/dt$), and using standard thermodynamic identities for the first partial derivatives of $S$ with respect to these variables, $dS/dt$ can be expressed as $dS/dt = (C_p/T)\, dT/dt + sI_t$. A term $(\alpha_4 V)\, dP/dt$, where $\alpha_4$ is the $^4$He thermal expansion coefficient, contributes negligibly and has been dropped. Equating the two above expressions for $dS/dt$ gives rise to equation 4. This is the same result as Eq. 7 of [14], in the appropriate limits.

**Appendix B -- Calibrations**

The diaphragm displacement transducer output signal is a voltage, $\Delta U$, proportional to displacement, $x = \alpha \Delta U$. Displacement is proportional to the pressure difference: $kx = \Delta PA$, where $k$ is the diaphragm spring constant and $A$ its area. Thus $\Delta P = k\alpha \Delta U/A \equiv \gamma \Delta U$. The constant $\gamma$ is determined from the Josephson frequency (Eq. 2) measured at the beginning of a pressure driven transient where $\Delta T = 0$ [5]. The total current is $I_t = \rho A\, \partial x/\partial t = \rho(A^2/k)\, \partial \Delta P/\partial t$. With constant heater power $W_h$, any thermally driven transient will eventually reach a steady state where $\Delta\mu = 0$, $\partial\Delta T/\partial t = 0$, and $I_s = -I_n$. In this case, equations 4 and 5 give $\Delta T = \Delta P_f/\rho s = RW_h/(1+\rho^2 s^2 TR\beta/\eta)$, from which the thermal resistance $R$ is determined for a given $T$ from the measured $\Delta P_f$ versus $W_h$. If the heater power is sufficiently small, no Josephson oscillations are excited and $\Delta\mu = 0$ is maintained throughout the transient. In this case $\Delta P_f$ and $\Delta T$ will decay exponentially with time constant $\tau = R(C_p + s^2\rho^2 TA^2/k)/(1+\rho^2 s^2 TR\beta/\eta)$, as can be shown from equations 4 and 5 using $\Delta P_f = \rho s \Delta T$ and $I_s = \rho(A^2/k)d\Delta P/dt - I_n$. The constants $A^2/k$ and $V$ (recall $C_p = c_p V$) are determined from a fit to the measured $\tau$ as a function of $T$. The parameter values, with uncertainty in the last digit specified in brackets, are: $\gamma = 0.0313(6)\,\text{Pa/V}$, $R = 17.5(5)\,\text{K/W}$, $\beta = 4.8(3)\times 10^{-20}\,\text{m}^3$, $V = 2.45(5)\times 10^{-8}\,\text{m}^3$, $k/A^2 = 1.88(4)\times 10^{12}\,\text{N/m}^5$. The power applied to the heater for the transient shown in figures 2 and 3 was $W_h = 40.4(4)\,\text{nW}$.

Prior to the above calibrations, the inner cell volume $V$ was known to within 20%. If this value is used for $V$ (instead of having $V$ be a fit parameter), the pressure calibration constant $\gamma$ can be determined independently of any Josephson frequency measurement as follows. For constant heater power $W_h$, we have (in steady state) $RW_h/(1+\rho^2 s^2 TR\beta/\eta) = \gamma\Delta U_f/\rho s$, where $\Delta U_f = \Delta P_f/\gamma$ is measured. If this expression is substituted into the equation for $\tau$, then a fit to $\tau$ as a function of $T$ gives $A^2/k$ and $\gamma$. This procedure was used in the initial demonstration of pressure driven Josephson oscillations [5].

**Acknowledgements**

We would like to thank Dr. K. Penanen for suggesting searching for the temperature drive phenomenon by the inclusion of a heater in our conventional cells. Dr. T. Haard made important contributions to the construction and design of the apparatus. M. Abreau also helped with assembly of the apparatus. The aperture arrays were fabricated by A. Loshak. This work was supported in part by the NSF (grant DMR 0244882) and NASA.

1. P. W. Anderson, Rev. Mod. Phys. **38**, 298 (1966).

2. R. E. Packard, Rev. Mod. Phys. **70**, 641 (1998).

3. L. D. Landau and E. M. Lifshitz, *Statistical Physics, 3rd ed.* (Butterworth-Heinemann, Oxford, 1980), Eq. 24.12.

4. S. V. Pereversev, A. Loshak, S. Backhaus, J. C. Davis, and R. E. Packard, Nature **388**, 449 (1997).

5. E. Hoskinson, T. M. Haard, and R. E. Packard, Nature **433**, 376 (2005).

6. H. J. Paik, J. Appl. Phys. **47**, 1168 (1976).

7. D. R. Tilley and J. Tilley, *Superfluidity and Superconductivity, 3$^{rd}$ ed.* (Institute of Physics, Bristol and Philadelphia, 1990), Eq. 3.47.

8. R. J. Donnelly and C. F. Barenghi, J. Phys. Chem. Ref. Data. **27**, 1217 (1998).

9. J. C. Davis, R. E. Packard, Rev. Mod. Phys. **74**, 741 (2002).

10. K. Sukhatme, Y. Mukharsky, T. Chui, and D. Pearson, Nature **411**, 280 (2001).

11. E. Varoquaux, M. Meisel, and O. Avenel, Phys. Rev. Lett. **57**, 2291 (1986).

12. K. Penanen, APS March meeting, L15.007 (2004).

13. O. Avenel, Yu. Mukharsky, and E. Varoquaux, J. Low. Temp. Phys. **135**, 745 (2004).

14. S. Backhaus and E. Yu. Backhaus, J. Low. Temp. Phys. **109**, 511 (1997).